\documentclass[aps,prd,onecolumn,groupedaddress,showpacs,nofootinbib,amssymb]{revtex4}
\usepackage[T1]{fontenc}
\usepackage[latin1]{inputenc}
\usepackage{graphicx,color}
\usepackage[english]{babel}
\usepackage{amsmath}
\usepackage{amssymb}
\usepackage{amsfonts}

\begin{document}

\newcommand{\be}{\begin{equation}}
\newcommand{\ee}{\end{equation}}
\newcommand{\bea}{\begin{eqnarray}}
\newcommand{\eea}{\end{eqnarray}}
\newcommand{\beaa}{\begin{eqnarray*}}
\newcommand{\eeaa}{\end{eqnarray*}}
\newcommand{\Lhat}{\widehat{\mathcal{L}}}
\newcommand{\nn}{\nonumber \\}
\newcommand{\e}{\mathrm{e}}
\newcommand{\tr}{\mathrm{tr}\,}

\tolerance=5000

\title{Models for Little Rip Dark Energy}
\author{Paul H. Frampton$^{1}$\footnote{frampton@physics.unc.edu}, 
Kevin J. Ludwick$^{1}$\footnote{kludwick@physics.unc.edu},
Shin'ichi Nojiri$^2$\footnote{nojiri@phys.nagoya-u.ac.jp},
Sergei D. Odintsov$^{2,3}$\footnote{odintsov@aliga.ieec.uab.es}$^{,}$\footnote{Also at Tomsk 
State Pedagogical University.},
and Robert J. Scherrer,$^{4}$\footnote{robert.scherrer@vanderbilt.edu}}
\affiliation{$^1$Department of Physics \& Astronomy, University of North
Carolina, Chapel Hill, NC~~27599}
\affiliation{$^2$Department of Physics, Nagoya University, Nagoya 464-8602, Japan \\
and Kobayashi-Maskawa Institute for the Origin of Particles and the Universe,
Nagoya University, Nagoya 464-8602, Japan}
\affiliation{$^3$Instituci\`{o} Catalana de Recerca i Estudis Avan\c{c}ats (ICREA)
and Institut de Ciencies de l'Espai (IEEC-CSIC),
Campus UAB, Facultat de Ciencies, Torre C5-Par-2a pl, E-08193 Bellaterra
(Barcelona), Spain}
\affiliation{$^4$Department of Physics \& Astronomy, Vanderbilt University,
Nashville, TN~~37235}
\date{\today}

\begin{abstract}

We examine in more detail specific models which yield a little rip cosmology,
i.e., a universe in which the dark energy density increases without bound but
the universe never reaches a finite-time singularity. We derive the conditions
for the little rip in terms of the inertial force in the expanding universe
and present two representative models to illustrate in more detail
the difference between little rip models and those which are asymptotically
de Sitter. We derive conditions on the equation of state parameter of
the dark energy to distinguish between the two types of models.
We show that coupling between dark matter and dark energy with a little rip
equation of state can alter the evolution, changing the little rip into
an asymptotic de Sitter expansion. We give conditions on minimally-coupled
phantom scalar field models and on scalar-tensor models that indicate whether or
not they correspond to a little rip expansion.  We show that, counterintuitively, 
despite local instability, a little-rip cosmology has an infinite lifetime.

\end{abstract}


\maketitle

\section{Introduction}

The current acceleration of the universe is often attributed to dark energy, an 
unknown fluid with effective equation of state (EoS) parameter $w$ close to 
$-1$. The observational data \cite{obs} favor
$\Lambda$CDM with $w=-1$. However, phantom ($w<-1$) or 
quintessence ($-1/3> w > -1$) 
dark energy models are not excluded by observational data \cite{Li:2011sd}. In both 
cases, it is known that the universe may evolve to a finite-time future 
singularity. 
Phantom dark energy models can lead to a singularity in which the scale
factor and density become
infinite at a finite time;
such a singularity is called 
a big rip \cite{Frampton:2003,Caldwell:2003vq}, or Type I singularity
\cite{Nojiri:2005sx}. 
For quintessence dark energy, one can have a singularity
for which the pressure goes to infinity
at a fixed time, but the scale factor and density remain finite;
this is called
a sudden singularity \cite{Barrow:2004xh,Barrow:2004hk}, or a Type II singularity
\cite{Nojiri:2005sx}). Alternately, the density and pressure can both become infinite
with a finite scale factor at a finite time
(a Type III singularity), or higher derivatives of the Hubble parameter $H$
can diverge (a Type IV singularity) \cite{Nojiri:2005sx}.
The occurrence of a singularity at a finite time in the future 
may lead to some inconsistencies.
Several scenarios to avoid a future singularity have been proposed so far: 
coupling with dark matter \cite{Nojiri:2009pf}, 
inclusion of quantum effects \cite{Elizalde:2004mq}, 
additional changes in the equation of state \cite{g} or special forms of 
modified gravity \cite{g}.

Recently, a new scenario to avoid a future singularity has been proposed in 
Ref.~\cite{Frampton:2011sp}. In this scenario, $w$ is less than $-1$, so that
the dark energy density increases with time, but
$w$ approaches $-1$ asymptotically and sufficiently rapidly that
a singularity is avoided. This proposed non-singular cosmology was called a ``little rip'' 
because it leads to a dissolution of bound structures at some point in the future 
(similar to the effect of a big rip singularity). It can be realized in terms of a general 
fluid with a complicated EoS \cite{Nojiri:2005sx,pr}. 
The evolution of the little rip cosmology is close to that of $\Lambda$CDM up to the present, 
and is
similarly consistent with the observational data.

The present article is devoted to further study of the properties of the little rip cosmology. 
In the next section, the inertial force interpretation of the little rip is developed,
and it becomes clear why a 
dissolution of bound structures occurs. Coupling of the little rip fluid with dark matter 
is considered in Section III. It is shown that as the result of such a coupling an asymptotically 
de Sitter universe can eventually evolve to have a little or big rip. In Section IV, the
little rip cosmology is reconstructed in terms of scalar field models.
Our results are summarized
in Section V.

\section{Inertial force interpretation of the little rip}

As the universe expands, the relative acceleration between two points separated by a comoving 
distance $l$ is given by
$l \ddot a/a$, where $a$ is the scale factor. 
An observer a comoving distance $l$ away from a mass $m$ will measure an inertial force on the mass of
\be
\label{i1}
F_\mathrm{iner}=m l \ddot a/a = m l \left( \dot H + H^2 \right)\, .
\ee 
Let us assume the %the 
two particles are bound by a constant force $F_0$. If $F_\mathrm{iner}$ is positive and greater 
than $F_0$, the two particles become unbound. This is the ``rip'' produced by the accelerating 
expansion.
%%%%%%%%%%%%%%%%%%%%%
%\footnote{
%As we will see later, we cannot find if the little rip could occur or not (by the comparison of (\ref{i7}) 
%with (\ref{i4})). We can, however, find it by the $a$-dependence of $w$ as clear from  (\ref{wa}). }
%%%%%%%%%%%%%%%%%%%%%
Note that equation (\ref{i1}) shows that a rip always occurs when either $H$ diverges or $\dot H$
diverges (assuming $\dot H > 0$). The first case corresponds to 
a ``big rip'' \cite{Caldwell:1999ew}, while
if $H$ is finite, but $\dot H$ diverges with $\dot H > 0$,
we have a Type II or ``sudden future'' singularity 
\cite{Barrow:2004xh,Barrow:2004hk,Nojiri:2005sx}, which also leads to a rip. 
However, as noted in Ref. \cite{Frampton:2011sp}, it is possible for $H$,
and therefore, 
$F_\mathrm{iner}$, to increase without bound and yet not produce a future
singularity at a finite time; this is the little rip.  Both the big rip and
little rip are characterized by $F_\mathrm{iner} \rightarrow \infty$; the difference
is that $F_\mathrm{iner} \rightarrow \infty$ occurs at a finite time for a big rip and as $t \rightarrow \infty$
for the little rip.

An interesting case occurs when $H$ is finite and $\dot H$ diverges but is negative. In this case, 
even though the universe is expanding, all structures are crushed rather than ripped. An example is given by
\be
\label{lc1}
H = H_0 + H_1 \left(t_c - t\right)^\alpha\, .
\ee
Here $H_0$ and $H_1$ are positive constants and $\alpha$ is a constant with $0<\alpha<1$.

By using the FRW equations
\be
\label{FRW}
\frac{3}{\kappa^2}H^2=\rho\, ,\quad 
 - \frac{1}{\kappa^2}\left( 2\dot H + 3 H^2 \right) = p\, ,
\ee
we may rewrite (\ref{i1}) in the following form:
\be
\label{i1b}
F_\mathrm{iner}= - \frac{ml\kappa^2}{6} \left( \rho + 3p \right)\, .
\ee
Here $\kappa^2 = 8\pi G$ and $G$ is Newton's gravitational constant. 
Not surprisingly, we see that the inertial force is sourced by the quantity $\rho + 3p$.
Then if we consider the general equation of state,
\be
\label{gEoS}
p = - \rho + f(\rho)\, ,
\ee
we find 
\be
\label{i1c}
F_\mathrm{iner}= \frac{ml\kappa^2}{6} \left( 2 \rho - 3f(\rho) \right)\, .
\ee

As noted in Ref.~\cite{Frampton:2011sp}, when $w\to -1$ but $w<-1$, a rip can
occur without a singularity.
If we ignore the contribution from matter, the equation of state (EoS) parameter $w$ of the dark energy can be 
expressed in terms of the Hubble rate $H$ as
\be
\label{i2}
w= - 1 - \frac{2\dot H}{3H^2}\, .
\ee
Then if $\dot H>0$, we find $w<-1$. 

Now consider the following example:
\be
\label{i3}
H = H_0 \e^{\lambda t}\, .
\ee
Here $H_0$ and $\lambda$ are positive constants. 
Eq.~(\ref{i3}) tells us that there is no curvature singularity for finite $t$. 
By using Eq.~(\ref{i2}), we find
\be
\label{i4}
w = -1 - \frac{2\lambda}{3H_0} \e^{-\lambda t}\, ,
\ee
and therefore $w<-1$ and $w\to -1$ when $t\to +\infty$, and 
$w$ is always less than $-1$ when $\dot H$ is positive. 
 From Eq.~(\ref{i1}), we have
\be
\label{i5}
F_\mathrm{iner}=m l \left(\lambda H_0 \e^{\lambda t} + H_0^2 \e^{2\lambda t} \right)\, ,
\ee
which is positive and unbounded. Thus, $F_\mathrm{iner}$
becomes arbitrarily large with increasing $t$, resulting in a little rip. 

As another example, consider the model:
\be
\label{i6}
H = H_0 - H_1 \e^{-\lambda t}\, .
\ee
Here $H_0$, $H_1$, and $\lambda$ are positive constants and we assume $H_0>H_1$ and $t>0$. 
Since the second term decreases when $t$ increases, the universe goes to asymptotically 
de Sitter space-time. 
Then from Eq.~(\ref{i2}), we find
\be
\label{i7}
w = -1 - \frac{2\lambda H_1 \e^{- \lambda t}}{3\left( H_0 - H_1 \e^{-\lambda t} \right)^2} \, .
\ee
As in the previous example, $w<-1$ and $w\to -1$ when $t\to +\infty$. 
For $H$ given by Eq.~(\ref{i6}), however, the inertial force, given by (\ref{i1}), is
\be
\label{i8}
F_\mathrm{iner}
=m l \left\{ \lambda H_1 \e^{-\lambda t} + \left(H_0 - H_1 \e^{-\lambda t} \right)^2 \right\}\, ,
\ee
which is positive but bounded and $F_\mathrm{iner} \to m l H_0^2$ when $t\to +\infty$. 
Therefore if we choose $H_0$, $H_1$, and $\lambda$ small enough, we do not
obtain a rip. 
When $t$ becomes large, the scale factor $a$ is given by that of 
the de Sitter space-time $a\sim a_0\e^{H_0 t}$, and the energy density $\rho$ has the following 
form:
\be
\label{rho1}
\rho = \frac{3}{\kappa^2}H^2 \sim \frac{3}{\kappa^2}\left( H_0^2 - 2 H_0 H_1 \e^{-\lambda t} \right) 
\sim \frac{3}{\kappa^2}\left( H_0^2 - 2 H_0 H_1 \left(\frac{a}{a_0}\right)^{-\frac{\lambda}{H_0}} \right)\, ,
\ee
which is an increasing function of $a$ and becomes finite as $a \rightarrow \infty$. 

For $t \rightarrow \infty$, Eq.~(\ref{i7}) gives the asymptotic behavior of $w$ to be
\be
\label{wn1}
w \sim -1 - \frac{2\lambda H_1 \e^{- \lambda t}}{3H_0^2} \, ,
\ee
which is identical with (\ref{i4}) if we replace ${\lambda H_1}/{H_0}$
with $\lambda$.

These results indicate that knowledge of the asymptotic ($t \rightarrow \infty$) behavior
of $w(t)$ is insufficient to distinguish models with a rip from models which are asymptotically
de Sitter. The reason for this becomes clear when we derive the expression for $\rho(t)$
as a function of $w(t)$. The evolution of $\rho$ is given by:
\be
\label{drhodt}
\frac{d\rho}{dt} = -3H(\rho + p)\, ,
\ee
which can be expressed as
\be
\rho^{-3/2} \frac{d\rho}{dt} = -\sqrt{3} \kappa (1+w) \, .
\ee
Integrating between initial and final times $t_i$ and $t_f$ gives:
\be
\rho_i^{-1/2} - \rho_f^{-1/2} = -\frac{\sqrt{3}}{2}\int_{t_i}^{t_f} [1+w(t)]dt\, .
\ee
Evolution leading to a little rip implies that $\rho_f \rightarrow \infty$ as $t_f \rightarrow \infty$,
while asymptotic de Sitter evolution requires $\rho_f \rightarrow \mbox{constant}$
as $t_f \rightarrow \infty$. However, in either case, the integral on the right-hand side
simply approaches a constant as the upper limit goes to infinity. Thus, the
asymptotic functional
form for $w(t)$ is not a good test of the asymptotic behavior of $\rho$.

On the other hand, expressing the equation of state parameter as a function
of the scale factor $a$ instead of the time $t$ does provide a clearer test of the
existence of a future rip. Equation (\ref{drhodt}) can be written in terms
of the scale factor as
\be
\frac{a}{\rho}\frac{d\rho}{da} = -3[1+w(a)]\, ,
\ee
from which it follows that
\be
\ln\left(\frac{\rho_f}{\rho_i}\right) = -3 \int_{a_i}^{a_f} [1+w(a)]\frac{da}{a}\, .
\label{wa}
\ee
Thus, $\rho$ is asymptotically constant if the integral of
$(1+w)/a$ converges at its upper limit, while $\rho$ will increase without bound, leading to a rip,
when the integral diverges.
%%%%%%%%%%%%%%%
Then if $1+w(a)$ behaves as an inverse power of $a$, as in $1+w(a) \sim a^{-\epsilon}$ with 
arbitrary positive constant $\epsilon$ when $a\to \infty$, the integration on
the right-hand side 
of (\ref{wa}) is finite when $a_f\to \infty$, and therefore a rip does not
occur.
If $1+w(a)$ vanishes more slowly than any power of $a$ when $a\to \infty$, e.g., 
$1+w(a) \sim 1/\ln a$, the integration on the right-hand side of (\ref{wa}) diverges when $a_f\to \infty$, 
and therefore a rip is generated. 
%%%%%%%%%%%%%%%

We now consider what kind of perfect fluid realizes the evolution of $H$ in 
Eqs.~(\ref{i3}) or (\ref{i6}). 
The FRW equations give
\be
\label{i9}
\rho = \frac{3}{\kappa^2}H^2\, ,\quad \rho + p = - \frac{2}{\kappa^2} \dot H \, .
\ee
Consider first the model given by Eq.~(\ref{i3}).
By substituting Eq.~(\ref{i3}) into Eq.~(\ref{i9}) and eliminating
$t$, we obtain:
\be
\label{i10}
\left( \rho + p \right)^2 = \frac{4\lambda^2}{3\kappa^2} \rho\, .
\ee
On the other hand, for the case corresponding to Eq.~(\ref{i6}), we obtain:
\be
\label{i11}
\rho = \frac{3H_0^2}{\kappa^2} + \frac{3H_0}{\lambda}\left(\rho + p \right) 
+ \frac{3\kappa^2}{4\lambda^2} \left( \rho + p \right)^2\, .
\ee

\section{Coupling with dark matter \label{DEDM}}

In Ref.~\cite{Nojiri:2009pf}, it was shown that the coupling of
zero-pressure dark matter with 
phantom dark %matter 
energy
%%%%%%
could avoid a big rip singularity, and the universe might
evolve to asymptotic de Sitter 
space. Here we investigate the possibility that
coupling with the dark matter could avoid a little rip. 
We consider the equation of state Eq. (\ref{i10}), for which a little rip occurs in
the absence of such a coupling. 
We show that by adding a coupling with dark matter, a little rip can be avoided, and
the universe can evolve to de Sitter space.

We now consider the following conservation law \cite{Nojiri:2009pf}
\be
\label{dm1}
\dot \rho + 3H \left( \rho + p \right) = - Q \rho\, ,\quad 
\dot \rho_\mathrm{DM} + 3 H \rho_\mathrm{DM} = Q \rho\, .
\ee
Here $\rho_\mathrm{DM}$ is the energy density of the dark matter and $Q$ is a positive constant. 
The right-hand sides in Eqs.~(\ref{dm1}) express the decay of the dark energy into dark matter. 
We assume the equation of state given in Eq.~(\ref{i10}), for which a rip could occur. 
Then the first equation in (\ref{dm1}) can be rewritten as
\be
\label{dm1b}
\dot\rho - \frac{2\lambda\sqrt{3\rho}}{\kappa} H = - Q\rho\, .
\ee
Note that $\rho + p < 0$ since we are considering the model $w<-1$. 

We now assume the de Sitter solution where $H$ is a constant: $H=H_0>0$. If we neglect the contribution from 
everything other than the dark energy and dark matter, the first FRW equation
\be
\label{dm2}
\frac{3}{\kappa^2} H^2 = \rho + \rho_\mathrm{DM}\, ,
\ee
indicates that $\rho + \rho_\mathrm{DM}$ is a constant. Then Eq.~(\ref{dm1}) becomes
\be
\label{dm3}
0 = 3H_0 \left( \rho + p + \rho_\mathrm{DM} \right)\, .
\ee
Since $H=H_0>0$, we find
\be
\label{dm4}
\rho_\mathrm{DM} = - \rho - p \, .
\ee
Note that the above equation (\ref{dm4}) can be obtained from the conservation law (\ref{dm1}) 
and the first FRW equation (\ref{dm2}) without using any equation of state. 

Now we assume the equation of state (\ref{i10}). Combining Eqs.~(\ref{i10}) and (\ref{dm4}), we get
\be
\label{dm5}
\rho= \frac{3\kappa^2}{4\lambda^2} \rho_\mathrm{DM}^2 \, .
\ee
Since $\rho + \rho_\mathrm{DM}$ is a constant, Eq.~(\ref{dm5}) implies that $\rho_\mathrm{DM}$ and 
therefore $\rho$ is a constant. Then the second equation in (\ref{dm1}) gives
\be
\label{dm6}
\rho_\mathrm{DM} = \frac{4H_0\lambda^2}{\kappa^2Q}\, ,
\ee
and therefore, from (\ref{dm5}), we find
\be
\label{dm7}
\rho = \frac{12 H_0^2 \lambda^2}{\kappa^2 Q^2}\, .
\ee
Then by using the FRW equation (\ref{dm2}), we find
\be
\label{dm8}
H_0 = \frac{4\lambda^2}{3Q \left( 1 - \frac{4\lambda^2}{Q^2}\right)}\, .
\ee
This requires 
\be
\label{dm9}
\frac{\lambda}{Q} < \frac{1}{2}\, .
\ee
By using (\ref{dm8}), we can rewrite (\ref{dm6}) and (\ref{dm7}) as
\be
\label{dm10}
\rho_\mathrm{DM} = \frac{16\lambda^4}{3\kappa^2 Q^2 \left( 1 - \frac{4\lambda^2}{Q^2}\right)}\, , \quad 
\rho = \frac{64\lambda^6}{3\kappa^2 Q^4 \left( 1 - \frac{4\lambda^2}{Q^2}\right)^2}\, .
\ee
Then we obtain
\be
\label{dm10b}
\frac{\rho_\mathrm{DM}}{\rho} = \frac{Q^2 \left( 1 - \frac{4\lambda^2}{Q^2}\right)}{4\lambda^2}\, .
\ee

At the present time, ${\rho_\mathrm{DM}}/{\rho} \sim {1}/{3}$, and the fact that this ratio
is of order unity today is called the coincidence problem. This observed ratio can be obtained in our model
when ${\lambda^2}/{Q^2}\sim {3}/{16}$.

%%%%%%%%%%%%
De Sitter space can be realized by the $a$-independent energy density. 
The energy density of the phantom dark energy increases by the expansion but it decreases by the 
decay into the dark matter. 
On the other hand, the energy density of the dark matter decreases by the expansion but it increases by the 
decay of the dark energy. In the above solution, the decay of the dark energy into the dark matter 
balances with the expansion of the universe, and the energy densities of both the dark energy and 
dark matter become constant. This mechanism is essentially identical to one found in \cite{Nojiri:2009pf}. 

%It is not clear the origins of the dark energy and the dark matter in our model but 
%the right-hand-sides in (\ref{dm1}) tell that the dark energy directly decays into the dark matter 
%as in the decay of the radio active matters. If the decay could occur by the collision of the two 
%dark energy ``particle'' for example, the  right-hand-sides in (\ref{dm1}) does not proportional 
%to $\rho$ but $\rho^2$. 
%%%%%%%%%%%%

If the solution corresponding to de Sitter space-time is an attractor, 
the universe becomes asymptotic de Sitter space-time and 
any rip might be avoided. 
In order to investigate if the de Sitter space-time is an attractor or not, we consider the perturbation from 
the de Sitter solution in (\ref{dm8}) and (\ref{dm10}):
\be
\label{dm11}
H = \frac{4\lambda^2}{3Q \left( 1 - \frac{4\lambda^2}{Q^2}\right)} + \delta H\, , \quad 
\rho_\mathrm{DM} = \frac{16\lambda^4}{3\kappa^2 Q^2 \left( 1 - \frac{4\lambda^2}{Q^2}\right)}
+ \delta \rho_\mathrm{DM}\, , \quad 
\rho = \frac{64\lambda^6}{3\kappa^2 Q^4 \left( 1 - \frac{4\lambda^2}{Q^2}\right)^2} + \delta \rho\, .
\ee
Then the first FRW equation (\ref{dm2}) gives
\be
\label{dm12}
\frac{8\lambda^2}{\kappa^2 Q \left( 1 - \frac{4\lambda^2}{Q^2}\right)} \delta H 
= \delta \rho + \delta_\mathrm{DM}\, .
\ee
The conservation laws (\ref{dm1}) and (\ref{dm1b}) give
\bea
\label{dm13}
\delta \dot \rho &=& \frac{16\lambda^4}{\kappa^2 Q^2 \left( 1 - \frac{4\lambda^2}{Q^2}\right)} \delta H 
 - \frac{Q}{2} \delta \rho \, , \nn
\delta \dot \rho_\mathrm{DM} &=& 
 - \frac{16\lambda^4}{\kappa^2 Q^2 \left( 1 - \frac{4\lambda^2}{Q^2}\right)} \delta H 
+ Q \delta \rho
 - \frac{4\lambda^2}{Q \left( 1 - \frac{4\lambda^2}{Q^2}\right)} \delta \rho_\mathrm{DM} \, .
\eea
By eliminating $\delta H$ in (\ref{dm13}) using (\ref{dm12}), we obtain
\be
\label{dm14}
\frac{d}{dt} 
\left( \begin{array}{c} \delta \rho \\ \delta \rho_\mathrm{DM} \end{array} \right) 
= \left( \begin{array}{cc} - \frac{Q}{2} \left( 1 - \frac{4\lambda^2}{Q^2} \right) & \frac{2\lambda^2}{Q} \\
Q \left( 1 - \frac{2\lambda^2}{Q^2} \right) & 
 - \frac{\frac{2\lambda^2}{Q} \left( 3 - \frac{4\lambda^2}{Q^2} \right)}{ 1 - \frac{4\lambda^2}{Q^2} } 
\end{array} \right) 
\left( \begin{array}{c} \delta \rho \\ \delta \rho_\mathrm{DM} \end{array} \right) \, .
\ee
In order for the de Sitter solution in (\ref{dm8}) and (\ref{dm10}) to be stable, 
all the eigenvalues of the matrix in (\ref{dm14}) 
should be negative, which requires the trace of the matrix to be negative and the determinant to be positive, 
giving
\be
\label{dm15}
 - \frac{Q}{2} \left( 1 - \frac{4\lambda^2}{Q^2} \right) 
 - \frac{\frac{2\lambda^2}{Q} 
\left( 3 - \frac{4\lambda^2}{Q^2} \right)}{ 1 - \frac{4\lambda^2}{Q^2} } < 0\, ,\quad 
\lambda^2 > 0 \, .
\ee
The second condition can be trivially satisfied, and the first condition is also satisfied as long as 
(\ref{dm9}) is satisfied. 
Therefore the de Sitter solution in (\ref{dm8}) and (\ref{dm10}) is stable and therefore an attractor. 
This tells us that the coupling of the dark matter with the dark energy as in (\ref{dm1}) eliminates
the little rip.

Thus, if the universe where the dark energy dominates is realized, the universe will expand as in (\ref{i3}). 
If there is an interaction as given in (\ref{dm1}), the dark energy decay into dark matter will
yield asymptotic de Sitter space-time corresponding to Eq.~(\ref{dm8}). 

\section{Scalar field little rip cosmology}

\subsection{Minimally Coupled Phantom Models}

First consider a minimally coupled phantom field $\phi$ which obeys the equation of motion
\be
\label{phiev}
\ddot{\phi} + 3 H \dot{\phi} - V^\prime(\phi) = 0\, ,
\ee
where the prime denotes the derivative with respect
to $\phi$.
A field evolving according to equation (\ref{phiev}) rolls
uphill in the potential.  In what follows, we assume a monotonically increasing
potential $V(\phi)$.  If this is not the case, then it is possible for the field
to become trapped in a local maximum of the potential, resulting in asymptotic
de Sitter evolution.

Kujat, Scherrer, and Sen \cite{Kujat} derived the conditions on $V(\phi)$ to avoid
a big rip, namely $V^\prime/V \rightarrow 0$ as $\phi \rightarrow \infty$, and
\be
\label{ripcondition}
\int \frac{\sqrt{V(\phi)}}{V^\prime(\phi)} d\phi \rightarrow \infty\, .
\ee
When these conditions are satisfied, $w$ approaches $-1$ sufficiently
rapidly that a big rip is avoided. 
%%%%%%%%%%%%%
%We should also note that if the potential has a local maximum, the scalar field will 
%be trapped there and the evolution of universe will end to asymptotically de Sitter 
%space.
%%%%%%%%%%%%%

We now extend this argument to determine the conditions necessary to avoid a little rip.
Clearly, we will have $\rho \rightarrow$ constant if $V(\phi)$ is
bounded from above, so that $V(\phi) \rightarrow V_0$ (where $V_0$ is a constant)
as $\phi \rightarrow \infty$. We can show that this is also a necessary condition.
Suppose that $V(\phi)$ is not bounded from above, so that $V(\phi) \rightarrow \infty$
as $\phi \rightarrow \infty$. Then the only way for the density of the scalar field to
remain bounded is if the field ``freezes'' at some fixed value $\phi_0$. However,
this is clearly impossible from equation (\ref{phiev}), since it would require
$\ddot{\phi} = \dot{\phi} = 0$ while $V^\prime(\phi) \ne 0$. Thus, boundedness of the potential
determines the boundary between little rip and asymptotic de Sitter evolution.
Phantom scalar field models with bounded potentials have been discussed
previously in Ref.~\cite{Elizalde:2004mq}.

\subsection{Scalar-Tensor Models}

Using the formulation in Ref.~\cite{Nojiri:2005pu}, 
we now consider what kind of scalar-tensor model, with an action given by
\be
\label{ma7}
S=\int d^4 x \sqrt{-g}\left\{
\frac{1}{2\kappa^2}R - \frac{1}{2}\omega(\phi)\partial_\mu \phi
\partial^\mu\phi - V(\phi) \right\}\, , 
\ee
can realize the evolution of $H$ given in Eqs.~(\ref{i3}) or (\ref{i6}). 
Here $\omega(\phi)$ and $V(\phi)$ are functions of the scalar field $\phi$.
Since the corresponding fluid is phantom with $w<-1$, the scalar field must be 
a ghost with a non-canonical kinetic term. 
If we consider the model where $\omega(\phi)$ and $V(\phi)$ are given by a single function
$f(\phi)$ as follows,
\be
\label{ma10}
\omega(\phi)=- \frac{2}{\kappa^2}f''(\phi)\, ,\quad
V(\phi)=\frac{1}{\kappa^2}\left(3f'(\phi)^2 + f''(\phi)\right)\, ,
\ee
the exact solution of the FRW equations has the following form:
\be
\label{ma11}
\phi=t\, ,\quad H=f'(t)\, .
\ee
Then for the model given by Eq.~(\ref{i3}), we find
\be
\label{i12}
\omega(\phi) = - \frac{2\lambda H_0}{\kappa^2} \e^{\lambda \phi}\, ,\quad 
V(\phi) = \frac{1}{\kappa^2} \left( 3 H_0^2 \e^{2\lambda \phi} + \lambda H_0 \e^{\lambda \phi} \right)\, .
\ee
Furthermore, if we redefine the scalar field $\phi$ to $\varphi$ by 
\be
\label{i13}
\varphi = \frac{2\e^{\frac{\lambda}{2}\phi}}{\kappa} \sqrt{\frac{2H_0}{\lambda}}\, ,
\ee
we find that the action (\ref{ma7}) has the following form:
\be
\label{i14}
S=\int d^4 x \sqrt{-g}\left\{
\frac{1}{2\kappa^2}R + \frac{1}{2} \partial_\mu \varphi \partial^\mu\varphi 
 - \frac{3 \lambda^2 \kappa^2}{64} \varphi^4 - \frac{\lambda^2 }{8} \varphi^2 
\right\}\, .
\ee
Note that in the action (\ref{i14}), $H_0$ does not appear. 
This is because the shift of $t$ in (\ref{i3}) effectively changes $H_0$. 
The parameter $A$ in \cite{Frampton:2011sp} corresponds to $2\lambda/\sqrt{3}$ in (\ref{i3}) 
and is bounded as $2.74\times 10^{-3}\, \mathrm{Gyr}^{-1} \leq A \leq 9.67\times 
10^{-3}\, \mathrm{Gyr}^{-1}$, or $2.37\times 10^{-3}\, \mathrm{Gyr}^{-1} \leq \lambda \leq 8.37\times 
10^{-3}\, \mathrm{Gyr}^{-1}$, by the results of the Supernova Cosmology Project \cite{Amanullah:2010vv}.
In \cite{Frampton:2011sp}, it was shown that the model defined by Eq.~(\ref{i3}) can
give
behavior of the distance modulus versus redshift almost
identical to that of $\Lambda$CDM, so this model can be made consistent with observational data. 

As in Ref.~\cite{Frampton:2011sp}, we can generalize the behavior of this model to 
\be
\label{g1}
H = H_0 \e^{C \e^{\lambda t}}\, .
\ee
Here $H_0$, $C$, and $\lambda$ are positive constants. 
Then we find 
\be
\label{g2}
\omega(\phi)=- \frac{2}{\kappa^2} H_0 C \lambda \e^{C \e^{\lambda \phi}}\e^{\lambda \phi} \, ,\quad
V(\phi)=\frac{1}{\kappa^2}\left(3 H_0^2 \e^{2C \e^{\lambda \phi}} 
+ H_0 C \lambda \e^{C \e^{\lambda \phi}}\e^{\lambda \phi} \right)\, .
\ee
If we redefine the scalar field $\phi$ to $\varphi$ by
\be
\label{g3}
\varphi = \frac{\sqrt{2 H_0 C \lambda }}{\kappa}\int d\phi \e^{\frac{C}{2} \e^{\lambda \phi}}
\e^{\frac{\lambda}{2} \phi}
= \frac{1}{\kappa} \sqrt{ \frac{8 H_0 C}{\lambda}} \int^{\e^{\frac{\lambda}{2} \phi}} dx 
\e^{\frac{C}{2} x^2} 
= \frac{2}{\kappa}\sqrt{\frac{H_0 \pi}{\lambda}} \mathrm{Erfi}\left[\sqrt{\frac{C}{2}}
\e^{\frac{\lambda}{2}\phi}\right]\, ,
\ee
we may obtain the action where the kinetic term of the scalar field $\varphi$ is 
$+ \frac{1}{2} \partial_\mu \varphi \partial^\mu \varphi$. 
In (\ref{g3}), $\mathrm{Erfi}[x] = \mathrm{Erf}[ix]/i$ with $i^2 = -1$, where $\mathrm{Erf}[x]$ is the error function. 
In \cite{Frampton:2011sp}, it was shown that the model given by Eq.~(\ref{g1}) can also be consistent 
with the observations.

As in Ref.~\cite{Frampton:2011sp}, we can easily find models which show more complicated behavior of $H$
such as
\be
\label{gg}
H = H_0 \e^{C_0 \e^{C_1 \e^{C_2 \e^{\lambda t}}}}\, .
\ee 

On the other hand, in the model given by Eq.~(\ref{i6}), we find 
\be
\label{i15}
\omega(\phi) = - \frac{2\lambda H_1}{\kappa^2} \e^{-\lambda \phi}\, ,\quad 
V(\phi) = \frac{1}{\kappa^2} \left\{ 3 \left( H_0 - H_1 \e^{-\lambda \phi} \right)^2 
+ \lambda H_1 \e^{- \lambda \phi} \right\}\, ,
\ee
and by the redefinition
\be
\label{i16}
\varphi = \frac{2\e^{-\frac{\lambda}{2}\phi}}{\kappa} \sqrt{\frac{2H_1}{\lambda}}\, ,
\ee
we find that the action (\ref{ma7}) has the following form:
\be
\label{i17}
S=\int d^4 x \sqrt{-g}\left\{
\frac{1}{2\kappa^2}R + \frac{1}{2} \partial_\mu \varphi \partial^\mu\varphi 
 - \frac{1}{\kappa^2} \left[ 3 \left( H_0 - \frac{\lambda \kappa^2}{8} \varphi^2 \right)^2 
+ \frac{\lambda^2 \kappa^2}{8} \varphi^2 \right] \right\}\, .
\ee
In the action given by Eq. (\ref{i17}), $H_1$ does not appear. This is because 
the shift of $t$ in (\ref{i6}) effectively changes $H_1$. 

Eq.~(\ref{i13}) shows that in the infinite future $t=\phi\to + \infty$, $\varphi$ also goes to 
infinity, that is, the scalar field climbs up the potential to infinity. 
This climbing up the potential makes the Hubble rate grow and generates a rip due to the inertial 
force (\ref{i1}). 
On the other hand, Eq.~(\ref{i16}) tells us that when $\phi \to +\infty$, $\varphi$ vanishes. 
Note that the potential in (\ref{i17}) is a double well potential similar to the potential of the 
Higgs field, and $\varphi=0$ corresponds to the local maximum of the potential. Therefore, in the model 
given by (\ref{i17}), the scalar field climbs up the potential and arrives at the local maximum after an infinite 
time. The behavior of the scalar field is different from that of the canonical scalar field, which 
usually rolls down the potential. This phenomenon of how the scalar field climbs up the potential occurs 
due to the non-canonical kinetic term. For the canonical scalar field $\varphi_c$, the field equation 
has the form of $\nabla_t^2 \varphi_c = - V'(\phi)$, but if the sign of the kinetic term is changed, 
we obtain $\nabla_t^2 \varphi_c = V'(\phi)$ for a non-canonical scalar field. That is, the sign of the 
``force'' is effectively changed. 

We now investigate the stability of the solution (\ref{ma11}) in the model given by Eqs.~(\ref{ma7}) and (\ref{ma10}) 
by considering the perturbation from the solution (\ref{ma11}):
\be
\label{pt1}
\phi=t + \delta \phi(t)\, ,\quad H=f'(t) + \delta h(t)\, .
\ee
By using the FRW equations
\be
\label{pt2}
\frac{3}{\kappa^2} H^2 = \frac{1}{2}\omega(\phi) {\dot \phi}^2 + V(\phi)\, ,\quad 
- \frac{1}{\kappa^2} \left( 2 \dot H + 3 H^2 \right) = \frac{1}{2}\omega(\phi) {\dot \phi}^2 - V(\phi)\, ,
\ee
we find
\be
\label{pt3}
\frac{d}{dt} \left( \begin{array}{c} \delta h \\ \delta \phi \end{array} \right) 
= \left( \begin{array}{cc}
 - 6 f'(t) & 6f'(t) f''(t) + f'''(t) \\
 - 3\frac{f'(t)}{f''(t)} & 3 f'(t) 
\end{array} \right)
\left( \begin{array}{c} \delta h \\ \delta \phi \end{array} \right) \, .
\ee
In order for the solution (\ref{ma11}) to be stable, all the eigenvalues of the matrix in (\ref{pt3}) 
should be negative, which requires the trace of the matrix to be negative and the determinant to be positive, 
giving
\be
\label{pt4}
 -3 f'(t) < 0\, ,\quad 3 \frac{f'(t)f'''(t)}{f''(t)} > 0\, .
\ee
The first condition is trivially satisfied in the expanding universe since $f'(t) = H>0$. 
If the universe is in the phantom phase, where $f''(t) = \dot H > 0$, the second condition 
reduces to $f'''(t) = \ddot H>0$. Then the model corresponding to (\ref{i6}) is unstable but 
the model corresponding to (\ref{i3}) is stable. There are no local maxima in the potential in 
(\ref{i14}), so one would expect the field to climb the potential well to infinity, generating a rip. 
In general, in a model which generates a big or little rip, $H$ goes to infinity, which requires $\ddot H>0$. 
Therefore in the scalar field model generating a big or little rip, the solution corresponding to the rip 
is stable, and models that are asymptotically de Sitter can eventually evolve to have a rip.

%{\color{red} 

\section{Including matter}

In the previous sections, we have neglected the contribution from matter except
for the dark matter in Sec. \ref{DEDM}. 
In this section, we now consider the affect of additional matter components. 
We assume each component has a constant EoS parameter $w^i_\mathrm{matter}$. Then the energy density 
and pressure contributed by all of these components can be expressed as
\be
\label{mat1}
\rho_\mathrm{matter} = \sum_i \rho^i_0 a^{-3 \left( 1 + w^i_\mathrm{matter} \right)}\, ,\quad 
p_\mathrm{matter} = \sum_i w_i \rho^i_0 a^{-3 \left( 1 + w^i_\mathrm{matter} \right)}\, .
\ee
Here the $\rho^i_0$'s are constants. 
Even including these additional matter components, we can construct the scalar-tensor model realizing 
the evolution of $H$ by, instead of (\ref{ma10}), 
\bea
\label{any5}
\omega(\phi) &=& - \frac{2}{\kappa^2}g''(\phi) - \sum_i 
\frac{w^i_\mathrm{matter} + 1}{2}\rho^i_0
a_0^{-3(1+w^i_\mathrm{matter})}
\e^{-3(1+w^i_\mathrm{matter})g(\phi)}\, ,\nn
V(\phi) &=& \frac{1}{\kappa^2}\left(3g'(\phi)^2 + g''(\phi)\right)
+ \sum_i \frac{w^i_\mathrm{matter} -1}{2}\rho^i_0
a_0^{-3(1+w^i_\mathrm{matter})} \e^{-3(1+w^i_\mathrm{matter})g(\phi)} \, .
\eea
Then the solution of the FRW equations (\ref{FRW}) is given by
\be
\label{any6}
\phi=t\, ,\quad H=g'(t)\, ,\quad
\left(a=a_0 \e^{g(t)}\right)\, .
\ee

We may consider the example of (\ref{i3}), which gives
\be
\label{mat2}
a(t) = a_0 \e^{\frac{H_0}{\lambda} \e^{\lambda t}}\, .
\ee
Then by using the FRW equations (\ref{FRW}), we find the EoS parameter $w_\mathrm{DE}$ 
corresponding to the dark energy is given by
\bea
\label{mat3}
w_\mathrm{DE} &=& \frac{ \frac{3}{\kappa^2} H^2 - \rho_\mathrm{matter}}
{ - \frac{1}{\kappa^2} \left(2\dot H + 3H^2\right) - p_\mathrm{matter}} \nn
&=& \frac{ \frac{3}{\kappa^2} H_0^2 \e^{2\lambda t} - \sum_i \rho^i_0 a_0^{-3 \left( 1 + w^i_\mathrm{matter} \right)}
\e^{- \frac{3 \left( 1 + w^i_\mathrm{matter} \right) H_0}{\lambda} \e^{\lambda t}}}
{ - \frac{1}{\kappa^2} \left(2\lambda H_0 \e^{\lambda t} + 3H_0^2 \e^{2\lambda t} \right)
 - \sum_i w^i_\mathrm{matter} \rho^i_0 a_0^{-3 \left( 1 + w^i_\mathrm{matter} \right)}
\e^{- \frac{3 \left( 1 + w^i_\mathrm{matter} \right) H_0}{\lambda} \e^{\lambda t}}}\, .
\eea
When $t$ becomes large, the contribution from the matter components decreases rapidly and $w_\mathrm{DE}$ in (\ref{mat3}) 
coincides with $w$ in (\ref{i4}). 
The density parameter $\Omega_\mathrm{DE}$ of the dark energy is also given by
\be
\label{mat4}
\Omega_\mathrm{DE} = \frac{ \frac{3}{\kappa^2} H^2 - \rho_\mathrm{matter}}{\frac{3}{\kappa^2} H^2 }
= 1 - \frac{\kappa^2}{3H_0^2} \sum_i \rho^i_0 a_0^{-3 \left( 1 + w^i_\mathrm{matter} \right)}
\e^{- \frac{3 \left( 1 + w^i_\mathrm{matter} \right) H_0}{\lambda} \e^{\lambda t} - 2\lambda t}\, ,
\ee
which rapidly goes to unity when $t$ becomes large. 
It would be interesting to consider the cosmological perturbation in the model including the contribution 
from these matter components. 

Let $t=0$ represent the present. We now assume the matter consists only of dust
with a vanishing
EoS parameter. Then we find $\rho_\mathrm{matter}=\rho_0a_0^{-3 }
\e^{- \frac{3 H_0}{\lambda}}$ and $p_\mathrm{matter} =0$. Since $\Omega_\mathrm{DE}=0.74$, we find 
$\rho_\mathrm{matter} = 0.26 \times \frac{3 H_0^2}{\kappa^2}$ by using (\ref{mat4}). 
Since $H_0$ is the Hubble parameter in the present universe, 
we find 
$H_0=7.24\times 10^{-2}\, \mathrm{Gyr}^{-1}\left( \thickapprox 70\,\mathrm{km}/\mathrm{s\,Mpc} \right)$. 
Since 
$2.37\times 10^{-3}\, \mathrm{Gyr}^{-1} \leq \lambda \leq 8.37\times 
10^{-3}\, \mathrm{Gyr}^{-1}$ (see below (\ref{i14})), by using (\ref{mat3}), we find
$-0.97<w_\mathrm{DE}<-0.72$, which could be consistent with the observed value 
$w_\mathrm{DE} = -0.972^{+0.061}_{-0.060}$.  

%}

\section{Discussion}

Little rip models provide an evolution for the universe
intermediate between asymptotic de Sitter expansion and models with a big rip
singularity. We have shown that the EoS parameter $w$ as a
function of time is a less useful diagnostic of such behavior than
is $w$ as a function of the scale factor. As for the case of big rip
singularities, a little rip can be avoided if the dark energy is coupled
to the dark matter so that energy flows from the dark energy to the dark
matter. Minimally coupled phantom scalar field models can lead to viable
little rip cosmologies. 
The models we investigated that yield little rip evolution 
turned out to be stable against small perturbations, and we found that 
big rip evolution is also consistent with the conditions for stability. For phantom 
field models, rip-like behavior is an attractor.

It is interesting that it was recently demonstrated that the little rip 
cosmology may be realized by a viscous fluid \cite{Brevik:2011mm}.
It turns out that the viscous little rip cosmology can also be stable.

Scalar little rip dark energy represents a natural alternative to the 
$\Lambda$CDM model, which also leads to a non-singular cosmology. It remains to
consider the coupling of such a model with matter and to 
confront its predictions with observations.

It is known \cite{CHT} that in a local frame with a flat background,
a classical field theory with $w<-1$ has a negative kinetic energy term,
and the corresponding quantum field theory has a tachyonic instability
and a vacuum decay lifetime which appears finite, although possibly greater than
the age of the universe. Our result shows that in the presence
of a rip, the space-time expansion is so fast that this tachyonic
instability does not have time to destabilize the global geometry
and shows, interestingly, that the extraordinary conditions
of a little rip can lead to an infinite lifetime.

\acknowledgements
P.H.F. and K.J.L. were
supported in part by the Department of Energy (DE-FG02-05ER41418). 
S.N. was supported in part by Global COE Program of Nagoya University (G07)
provided by the Ministry of Education, Culture, Sports, Science \& Technology (S.N.); the
JSPS Grant-in-Aid for Scientific Research (S) \# 22224003 and (C) \# 23540296 (S.N.). 
S.D.O. was supported by MICINN (Spain) projects FIS2006-02842 and FIS2010-15640, 
by CPAN Consolider Ingenio Project, 
AGAUR 2009SGR-994, and JSPS Visitor Program (Japan) S11135. 
R.J.S. was supported in part by the Department of
Energy (DE-FG05-85ER40226).


\begin{thebibliography}{}
\bibitem{obs} 
S.~Perlmutter {\it et al.}  [SNCP Collaboration],
Astrophys.\ J.\  {\bf 517}, 565 (1999) 
[arXiv:astro-ph/9812133]; \\ 
A.~G.~Riess {\it et al.}  [Supernova Search Team Collaboration],
Astron.\ J.\  {\bf 116}, 1009 (1998)
[arXiv:astro-ph/9805201]; \\
D.~N.~Spergel {\it et al.}  [WMAP Collaboration],
Astrophys.\ J.\ Suppl.\  {\bf 148}, 175 (2003) 
[arXiv:astro-ph/0302209]; \\ 
D.~N.~Spergel {\it et al.}  [WMAP Collaboration],
Astrophys.\ J.\ Suppl.\  {\bf 170}, 377 (2007) 
[arXiv:astro-ph/0603449]; \\
E.~Komatsu {\it et al.}  [WMAP Collaboration],
Astrophys.\ J.\ Suppl.\  {\bf 180}, 330 (2009) 
[arXiv:0803.0547 [astro-ph]]; \\
E.~Komatsu {\it et al.}  [WMAP Collaboration],
Astrophys.\ J.\ Suppl.\  {\bf 192}, 18 (2011)
[arXiv:1001.4538 [astro-ph.CO]].
\bibitem{Li:2011sd}
M.~Li, X.~-D.~Li, S.~Wang, and Y.~Wang,
[arXiv:1103.5870 [astro-ph.CO]].
\bibitem{Frampton:2003}
P.~H.~Frampton and T.~Takahashi,
Phys.\ Lett.\  B {\bf 557}, 135 (2003)
[arXiv:astro-ph/0211544].
\bibitem{Caldwell:2003vq}
R.~R.~Caldwell, M.~Kamionkowski, and N.~N.~Weinberg,
Phys.\ Rev.\ Lett.\  {\bf 91}, 071301 (2003) 
[arXiv:astro-ph/0302506].
\bibitem{Nojiri:2005sx}
S.~Nojiri, S.~D.~Odintsov and S.~Tsujikawa,
Phys.\ Rev.\  D {\bf 71}, 063004 (2005)
[arXiv:hep-th/0501025].
\bibitem{Barrow:2004xh}
J.~D.~Barrow,
Class.\ Quant.\ Grav.\  {\bf 21}, L79 (2004)
[arXiv:gr-qc/0403084].
\bibitem{Barrow:2004hk}
J.~D.~Barrow,
Class.\ Quant.\ Grav.\  {\bf 21}, 5619 (2004)
[arXiv:gr-qc/0409062].
\bibitem{Nojiri:2009pf}
S.~Nojiri and S.~D.~Odintsov,
Phys.\ Lett.\  B {\bf 686}, 44 (2010)
[arXiv:0911.2781 [hep-th]].
\bibitem{Elizalde:2004mq}
E.~Elizalde, S.~Nojiri and S.~D.~Odintsov,
Phys.\ Rev.\  D {\bf 70}, 043539 (2004)
[arXiv:hep-th/0405034].
\bibitem{g} %0807.2575; 1011.0544.
K.~Bamba, S.~Nojiri and S.~D.~Odintsov,
JCAP {\bf 0810}, 045 (2008)
[arXiv:0807.2575 [hep-th]]; \\
S.~Nojiri and S.~D.~Odintsov,
arXiv:1011.0544 [gr-qc].
\bibitem{Frampton:2011sp}
P.~H.~Frampton, K.~J.~Ludwick, R.~J.~Scherrer,
Phys.\ Rev.\  D {\bf 84}, 063003 (2011) 
[arXiv:1106.4996 [astro-ph.CO]].
\bibitem{pr} 
S.~Nojiri and S.~D.~Odintsov,
Phys.\ Rev.\  D {\bf 72}, 023003 (2005)
[arXiv:hep-th/0505215]; \\
H.~Stefancic,
Phys.\ Rev.\  D {\bf 71}, 124036 (2005)
[arXiv:astro-ph/0504518].
\bibitem{Caldwell:1999ew}
R.~R.~Caldwell,
Phys.\ Lett.\  B {\bf 545}, 23 (2002)
[arXiv:astro-ph/9908168].
\bibitem{Nojiri:2005pu}
S.~Nojiri and S.~D.~Odintsov,
Gen.\ Rel.\ Grav.\  {\bf 38}, 1285 (2006)
[arXiv:hep-th/0506212].
\bibitem{Amanullah:2010vv}
R.~Amanullah, C.~Lidman, D.~Rubin, G.~Aldering, P.~Astier, K.~Barbary, M.~S.~Burns, A.~Conley {\it et al.},
[arXiv:1004.1711 [astro-ph.CO]].
\bibitem{Kujat} 
J.~Kujat, R.~J.~Scherrer, A.~A.~Sen,
Phys.\ Rev.\  {\bf D74}, 083501 (2006) 
[arXiv:astro-ph/0606735].
\bibitem{Brevik:2011mm}
I.~Brevik, E.~Elizalde, S.~Nojiri, S.~D.~Odintsov,
[arXiv:1107.4642 [hep-th]].
\bibitem{CHT}
S.~M.~Carroll, M.~Hoffman and M.~Trodden,
Phys.\ Rev.\  D {\bf 68}, 023509 (2003)
[arXiv:astro-ph/0301273]; \\
J.~M.~Cline, S.~Jeon and G.~D.~Moore,
Phys.\ Rev.\  D {\bf 70}, 043543 (2004)
[arXiv:hep-ph/0311312].
\end{thebibliography}
\end{document}